\documentclass{article} 
\usepackage{amssymb}

\usepackage[figuresright]{rotating}

\usepackage{amsmath}
\title{High statistics measurements \\of pedestrian dynamics}

\begin{document}
\maketitle

\centerline{\scshape Alessandro Corbetta }
\medskip
{\footnotesize
\centerline{CASA- Centre for Analysis, Scientific computing and Applications,}
  \centerline{Department of Mathematics and Computer Science,
Eindhoven University of Technology,}
 \centerline{ P.O. Box 513, 5600 MB Eindhoven, The Netherlands,}
 \centerline{Department of Structural, Geotechnical and Building Engineering,} 
 \centerline{Politecnico di Torino, Corso Duca degli Abruzzi 24,
10126 Torino, Italy}   
} 

\medskip

\centerline{\scshape Luca Bruno}
\medskip
{\footnotesize
 \centerline{Department of Architecture and Design,}
\centerline{Politecnico di Torino, Viale Mattioli 39, 10125, Torino, Italy}

}

\medskip

\centerline{\scshape Adrian Muntean}
\medskip
{\footnotesize
 \centerline{CASA- Centre for Analysis, Scientific computing and Applications,}
  \centerline{ICMS - Institute for Complex Molecular Systems}
  \centerline{Department of Mathematics and Computer Science,
Eindhoven University of Technology,}
 \centerline{ P.O. Box 513, 5600 MB Eindhoven, The Netherlands,}

}

\medskip
\centerline{\scshape Federico Toschi}
\medskip
{\footnotesize
  \centerline{Department of Physics and Department of Mathematics and
    Computer Science,} 
\centerline{Eindhoven University of Technology, P.O. Box 513, 5600 MB Eindhoven, The Netherlands,}
  \centerline{CNR-IAC, Via dei Taurini 19, 00185 Rome, Italy,}

}

\bigskip

\begin{abstract}

Understanding the complex behavior of pedestrians walking in crowds is
a challenge for both science and technology. In particular, obtaining
reliable models for crowd dynamics, capable of exhibiting
qualitatively and quantitatively the observed emergent features of
pedestrian flows, may have a remarkable impact for matters as
security, comfort and structural serviceability.  Aiming at a
quantitative understanding of basic aspects of pedestrian dynamics,
extensive and high-accuracy measurements of pedestrian trajectories
have been performed. More than $100.000$ real-life, time-resolved
trajectories of people walking along a trafficked corridor in a
building of the Eindhoven University of Technology, The Netherlands,
have been recorded. A measurement strategy based on Microsoft
Kinect\texttrademark has been used; the trajectories of pedestrians
have been analyzed as ensemble data. The main result consists of a
statistical descriptions of pedestrian characteristic kinematic
quantities such as positions and fundamental diagrams, possibly
conditioned to local crowding status (e.g., one or more pedestrian(s)
walking, presence of co-flows and counter-flows).

\end{abstract}

\section{Introduction}

Understanding the behavior of pedestrians walking in crowds is a
complex challenge for both science and technology. Pedestrians have
been proved to walk reflecting collective behaviors, which result from
self-organized processes based on local interactions among individuals
(cf. e.g. \cite{helbing2009pedestrian, moussaid2009experimental,
  schadschneider2009empirical}). Obtaining reliable mathematical
models for the dynamics of crowds, capable of exhibiting qualitatively
and quantitatively these emergent features, may hence have a
remarkable impact for matters as security
(\cite{schadschneider2009evacuation, helbing2012crowd}), comfort and
structural serviceability (\cite{nakamura2006lateral,
  dallard2001london}).

At the present time the largest part of crowd dynamics models are
still missing a systematic experimental verification. This is probably
due to a combination of several factors: firstly, the large variability
in pedestrian crowds, flow conditions and geometries;  secondly, the
difficulty and partial inadequacy of measurement tools to resolve
pedestrian motion, which often call for well controlled laboratory
conditions to achieve higher reliability
(\cite{DBLP:journals/ijon/BoltesS13, kretz}).  The quality and
quantity of data are of course of paramount importance to validate
crowd dynamics models beyond the most basic aspects such as e.g. mean
velocity.

It is worth noticing that the use of laboratory conditions may fix
constraints on both the possible geometries as well as on the actual
``behavior'' of the participants. Moreover, it may impose a physical limitation
in the statistical exploration of human walking behavior because
inter-subject and intra-subject variabilities
(\cite{vzivanovic2007probability}) may be reduced or
neglected.

Our goal is to be able to develop tools capable to enable the
investigation of the dynamics of pedestrian crowds at extremely
high-statistics and with high-quality recordings in real-life
settings. It is common experience that complex systems can display
statistics that strongly deviate from Gaussian normal
distributions. This is for example the case of the dynamics of small
particle matter in turbulent flows (see \cite{toschi2009lagrangian}). From that field of
research we borrow both the techniques to accurately reconstruct
pedestrian trajectories (by means of software developed for Particle
Tracking Velocimetry, see \cite{OpenPTV}) as well as the statistical
observables that allow to statistically quantify the phenomenology of crowds and
thus make the quantitative comparison with model possible.

In this framework, we report on a few basic preliminary explorations
of the crowd dynamics measurements we recently conducted. In Section
\ref{Installation}, we describe our measurements set-up inspired by
the work by \cite{seer2012kinects} which allowed the \textit{in-vivo}
measurement of more than 100.000 pedestrians. Then we list a few
features of the data ensemble built in Section
\ref{Results}. Finally, we close the paper with the discussions in
Section \ref{Discussion}.

\section{The installation}\label{Installation}
\begin{figure}[t]\vspace*{4pt} 
\centerline{\includegraphics[scale=.48]{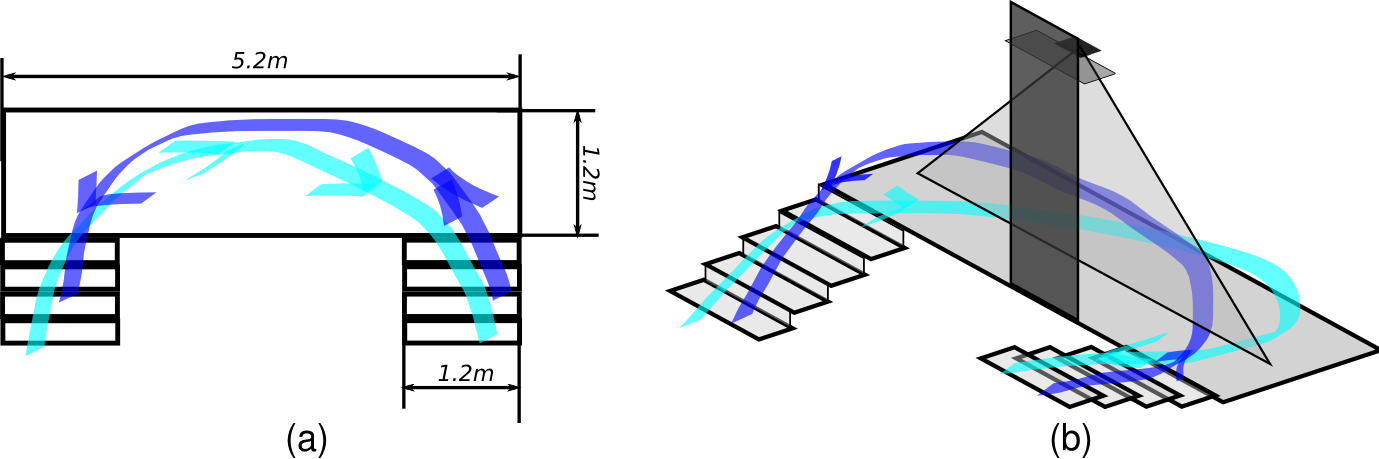}}
\caption{Schematics of the considered corridor. (a) top view; (b)
  three dimensional view. The bright arrow identifies the direction
  going from left to right ($2R$, ascending the stairs), the dark
  arrow otherwise ($2L$, descending the stairs).~\label{fig:corridor}}
\end{figure}

A relatively highly trafficked corridor at Eindhoven University of
Technology has been chosen as measurement site. The corridor connects
the canteen of the ``Metaforum'' building to its dining area and it is
crossed by approximately $2.200$ pedestrians every day. The observed
location serves as a landing between two different levels of the
building, hence it is globally U-shaped and has staircases at its
endings (see Fig.~\ref{fig:corridor} for a geometric reference). The
presence of the stairs induces a natural asymmetry between the two
span-wise walking actions: in particular, pedestrians going to the
right are \textit{ascending} the stairs, whilst pedestrians going to
the left are \textit{descending}. Even in such a simple, although
common, configuration many different walking scenarios might
occur. In the simplest case, pedestrians walk alone,
\textit{undisturbed} in their motion by the presence of peers. On the
other hand, when more than one person is present, either a
\textit{co-flow} or a \textit{counter-flow} condition might happen. In
the co-flow case, all pedestrians walk in the same direction, while,
in counter-flow case, a bi-directional flow occurs. It is important to
highlight that the data collected in this work do not refer to
pedestrians instructed \textit{a-priori} to cross the landing (as
common in many ``laboratory'' crowd experiments); rather, they refer
to the actual, unbiased, ``field'' measurement of pedestrian traffic.

The central, straight, section of the measurement site has been
recorded via a Microsoft Kinect\texttrademark special camera
(see~\cite{Kinect}) on a 24/7 basis and with a temporal resolution of
$15Hz$.  The Kinect\texttrademark is able to provide, on side of the
ordinary ``color'' picture of a target scene, that we disregarded, the
depth map: namely the distance map between every recorded pixel and
the camera plane (analogous to the one reconstructed
in~\cite{DBLP:journals/ijon/BoltesS13} by means of pairs of
cameras). This allows the development of pedestrian detection
algorithms with higher reliability.

The sensor has been located at an overhead altitude of $4m$, which
allows a recording window of span-wise length $3m$ (of which only $2.2m$
have been kept for accuracy reasons) and of width $1.2m$ (i.e., the
entire chord).

The sequence of depth maps provided by the sensor is streamed to a
processing unit which, as explained in the next Section, operates a
pedestrian head detection algorithm on a frame-by-frame basis and then
a multi-particle (head) tracking.

\subsection{Pedestrian detection and head tracking algorithms \label{sect:detection}}
The pedestrian detection algorithm is hereby concisely reported. For
more extensive insights, the reader can refer to the appendix
in~\cite{corbetta2014kinects} and to the original work
by~\cite{seer2012kinects}. The latter reference contains also
reliability estimates for the method.

Let $f^n=f^n(\vec{z})$ be the depth map recorded by
Kinect\texttrademark at time instant $n\geq 0$ and at spatial position
$\vec{z}=(x,y)$, i.e., in formulas,
\begin{equation}
f^n(\vec{z}):=distance(\mbox{element in $\vec{z}$} ,\mbox{camera
  plane}).
\end{equation}

\begin{figure}[t]\vspace*{4pt}
\centerline{\includegraphics[scale=.50]{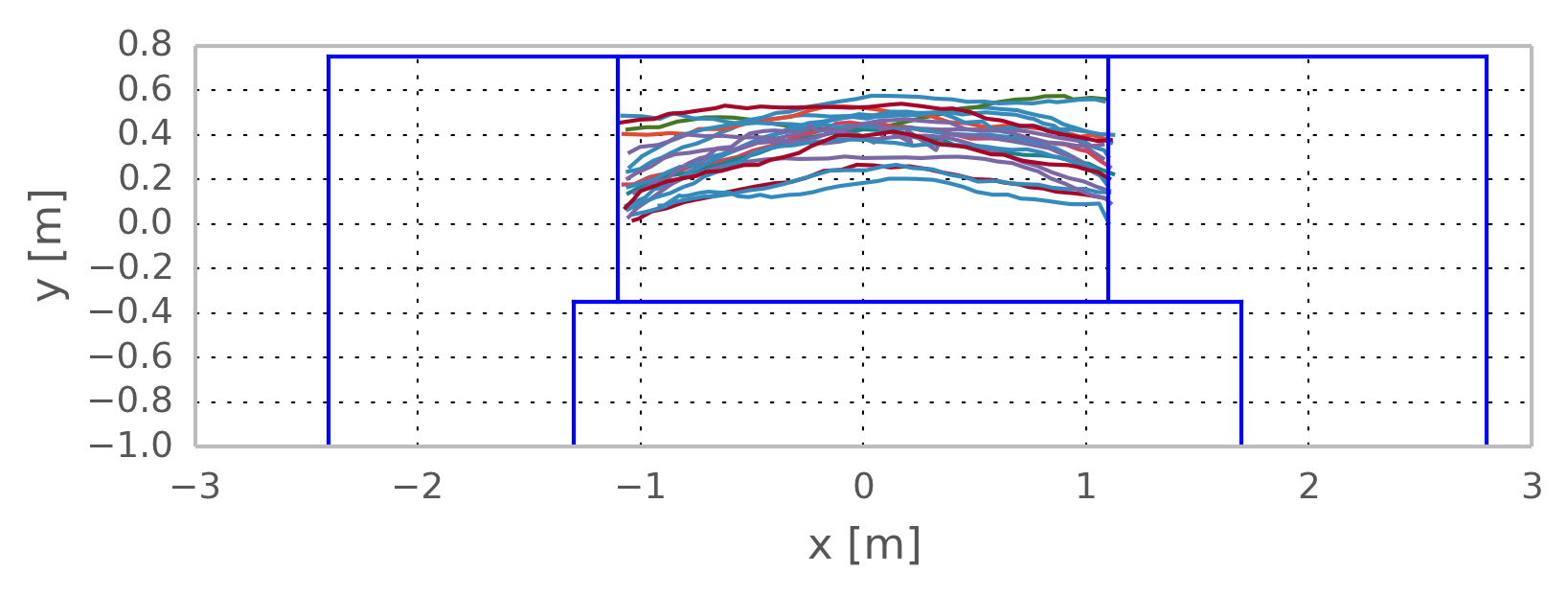}}%
\caption{Twenty trajectories chosen randomly among the ones recorded
  throughout the experimental campaign. These trajectories have been
  followed by pedestrians walking undisturbed from right to
  left (2L).~\label{fig:random-single-trajs}}
\end{figure}

To detect the positions of pedestrians in $f^n$ the following steps are performed.
\begin{enumerate}
\item \textbf{Depth-based significative foreground segmentation.} A
  common background $B=B(\vec{z})$ (possibly built after suitable
  averages of ``empty'' recordings) is expected across different depth
  maps. The foreground $\tilde F^n=\tilde F^n(\vec{z})$ is extracted
  via a thresholding operation
\[
\tilde F^n(\vec{z}) \leftarrow f^n(\vec{z}) \cdot [ f^n(\vec{z}) -
  B(\vec{z}) > \epsilon_1 ],
\]
where $\epsilon_1>0$ is a given (small) threshold, and $[P(\vec{z})] =
1$ whenever proposition $``P(\vec{z})"$ holds true, and $[P(\vec{z})]
= +\infty$ otherwise.

Besides, the foreground is likely populated by elements which are not
tall enough to be pedestrians, therefore a second thresholding
operation is performed
\[
F^n(\vec{z}) \leftarrow \tilde F^n(\vec{z}) \cdot [ \tilde F^n(\vec{z}) > h ].
\]
Here, $F^n = F^n(\vec{z})$ is the retained ``significative'' foreground and $h$ denotes the typical distance among the waist of pedestrians and the camera plane.
\item \textbf{Foreground random sampling.}  The ``dense'' foreground depth map $F^n$ is
  sampled and $N$ of its points (having finite depth value) are randomly extracted. A ``sparse'' (cloud) representative  of $F^n$,
\[
F^n_s = \{(\vec{z}_1,F^n(\vec{z}_1)),(\vec{z}_2,F^n(\vec{z}_2)),\ldots,(\vec{z}_N,F^n(\vec{z}_N))\},
\]
is hence built.

It is important to notice that the points left in $F^n_s$ likely belong to the pedestrians in the scene and, if $N$ is large enough ($N = O(500)$), they provide a good approximation of the three dimensional geometry of the latter. 

\item \textbf{Foreground random cloud clusterization.}  
To identify and isolate pedestrians, sparse samples in $F^n_s$ are
agglomerated in clusters which are likely in $1:1$ correspondence with
pedestrians themselves.

The agglomeration is performed via a \textit{hierarchical clustering}
operation based on the geometrical distance between points following a
\textit{maximum} linkage clustering (see,
e.g.,~\cite{duda2012pattern}).

Heuristically speaking, the sparse samples get iteratively
agglomerated in a binary tree fashion forming larger and larger
clusters. This iterative procedure merges clusters beginning from
single points on the basis of their distance; closest pairs are merged
first. Ideally, whenever a cluster $C^n$ features a distance from all
others clusters larger than the scale size $S$ of the human body, then
$C^n$ corresponds to a single pedestrian.

From a formal point of view, the length $S$ is adopted as cutoff
parameter of the clusterization tree, and the clusters
$C_1^n,C_2^n,\ldots,C_N^n$ underneath $S$ correspond to pedestrians
(see pedestrians in Fig.~\ref{fig:dendrogram}(a) which are identified
via the clusterization tree in Fig.~\ref{fig:dendrogram}(b)).

\begin{figure}[t]\vspace*{4pt}
\centerline{\includegraphics[scale=.65,trim=2cm 12.5cm 2cm
  2cm,clip]{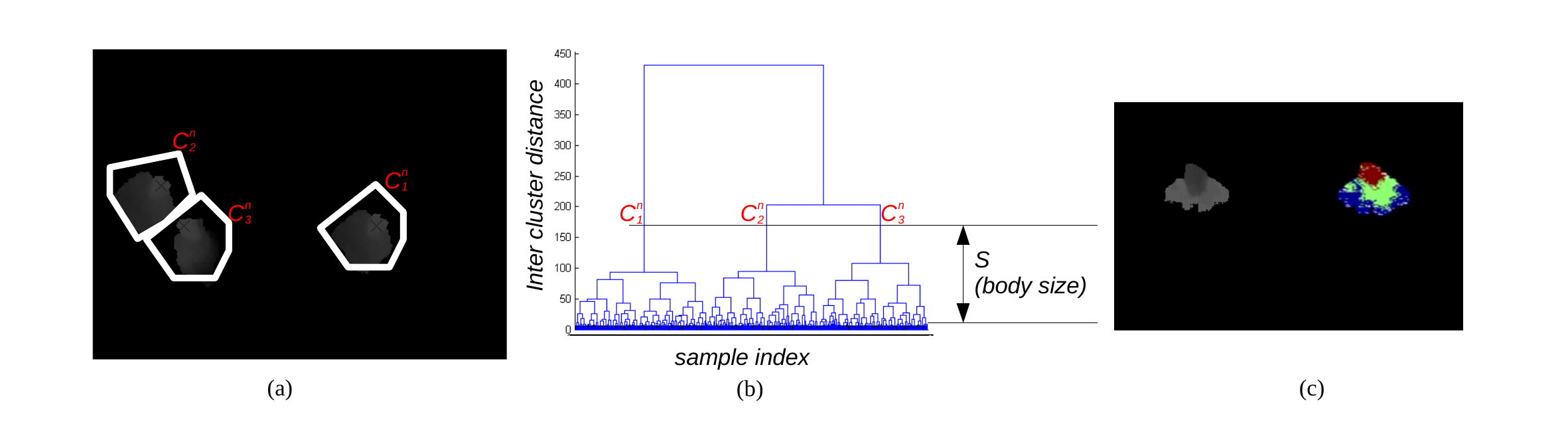}}%
\caption{(a) Foreground of a depth map showing three pedestrians; (b)
  clusterization tree of the random sampled foreground. The tree is
  cut at height $S$ and three different clusters, in correspondence
  with the pedestrians, are found; (c) (left) depth map of a single
  pedestrian as provided by Kinect; (right) random sampled version of
  the pedestrian. Colors identify region bounded by different depth
  percentiles: red (head), depth is less than the $10^{th}$
  percentile, green (shoulders), depth is between the $10^{th}$ and
  the $50^{th}$ percentile, blue (body), remaining
  points.~\label{fig:dendrogram}}
\end{figure}

\item \textbf{Head identification.}
Each point in a given cluster $C_i^n$ comes with a depth information;
therefore, the probability distribution function of the depths in
$C_i^n$ can be considered. The largest part of probability mass is
expected to be in the shoulder region, whilst the head marks a small
probability area having least distance from the camera. As a
consequence, the pedestrian head is identified as the set of points
$H_i^n \subset C_i^n$ such that they are closer to the camera than the
$10^{th}$ percentile of points in $C_i^n$ (see
Fig.~\ref{fig:dendrogram}(c)).

\item \textbf{Pedestrians tracking.}
Pedestrians are tracked following the centroid of their heads,
$\vec{Z}_i^n = mean(H_i^n)$, as they were particles. A tracking
approach analogous to Particle Tracking Velocimetry (PTV) from
experimental fluid mechanics is chosen (for a general reference on the
method, see e.g.~\cite{willneff2003spatio}. The actual tracking of
particles has been done via the OpenPTV library,
see~\cite{OpenPTV}). In Figure~\ref{fig:random-single-trajs}, we
illustrate a random selection of final trajectories obtained via this
method.
\end{enumerate}

\subsection{Ensemble data for pedestrian dynamics}
Once pedestrian trajectories have been reconstructed, they are
considered as elements of an ensemble
(see~\cite{corbetta2014kinects}). In other words, every pedestrian is
``confused'' in terms of its trajectory with all the others. The
analysis of such an ensemble allows one to isolate average behaviors
as well as ``unusual'' ones. This analysis can be enriched by looking
at specific flow conditions (e.g.~single pedestrians, counter-flows,
co-flows) at a time.  In this work, we focus on the average behaviors
intended as conditioned ensemble means. A study of the ``unusual''
behaviors will be considered in a future work.

\section{Results}\label{Results}
In this Section, we analyze the trajectories gathered from a
measurement campaign lasted $50$ working days, and which led to the
tracking of more than $100.000$ people. The trajectories spanned over
more than $2.5$ million depth frames in which up to $6$ people were
present at the same time (see Fig.~\ref{fig:load-distrib}(a).  In the
following Subsections different kinds of analyses grounded on
ensemble-means and depending on the local flow conditions are
performed. Specifically in~\ref{s:pedfeat} a quantitative overview on
the data is given; in~\ref{s:fdiag} fundamental diagrams obtained from
ensemble averages conditioned to the local pedestrian flow are
considered; finally, in~\ref{s:heat}, heat maps elaborated from
measured pedestrian positions are commented.

\subsection{Pedestrian data features~\label{s:pedfeat}}
\begin{figure}[t]\vspace*{4pt}
\centerline{\includegraphics[scale=.48]{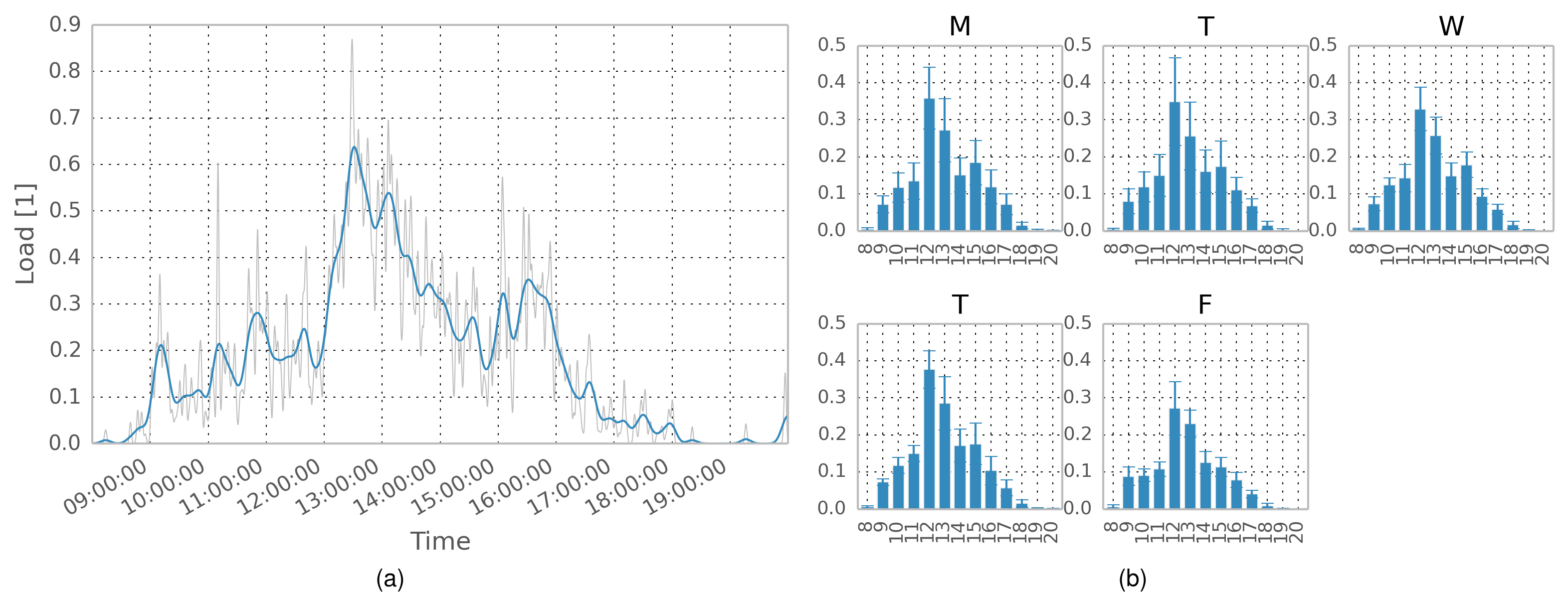}}%
\caption{(a) Time averaged facility load during the $19^{th}$ of June $2013$, from 8AM to 8PM  (width of the averaging windows: $1min$ - thin line, $5min$ - thick line); (b) facility load vs.~the hour of the day during the weekdays in terms of mean value (bars) and standard deviation (error bars).~\label{fig:daily-usage}} 
\end{figure}

The considered facility features different usage trends depending upon
the moment of the day. We consider the load, defined as the number of
pedestrians in the facility in a given instant of time, as primary
usage indicator. In formulas it reads as
\[
load(t) := \# \mbox{Pedestrians in the facility at time $t$}.
\]
In particular, we use time-averaged values of the load to easily
quantify the usage scenario as well as the most trafficked instants of
the day. In Fig.~\ref{fig:daily-usage}(a), we show a day-long time
history of the load. It shows two peaks: one at around 12PM (lunchtime
in Eindhoven) and another one at around 3PM (break). This usage trend
is homogeneous throughout the different weekdays as we report in
Fig.~\ref{fig:daily-usage}(b), where hourly averaged load statistics
depending on the day of the week are shown. From
Fig.~\ref{fig:daily-usage}(b) one may notice that the usage is
homogeneous across the weekdays with the exception of Fridays, in
which the traffic is generally reduced by approximately $20\%$.

\begin{figure}[t]\vspace*{4pt}
\centerline{\includegraphics[scale=.48]{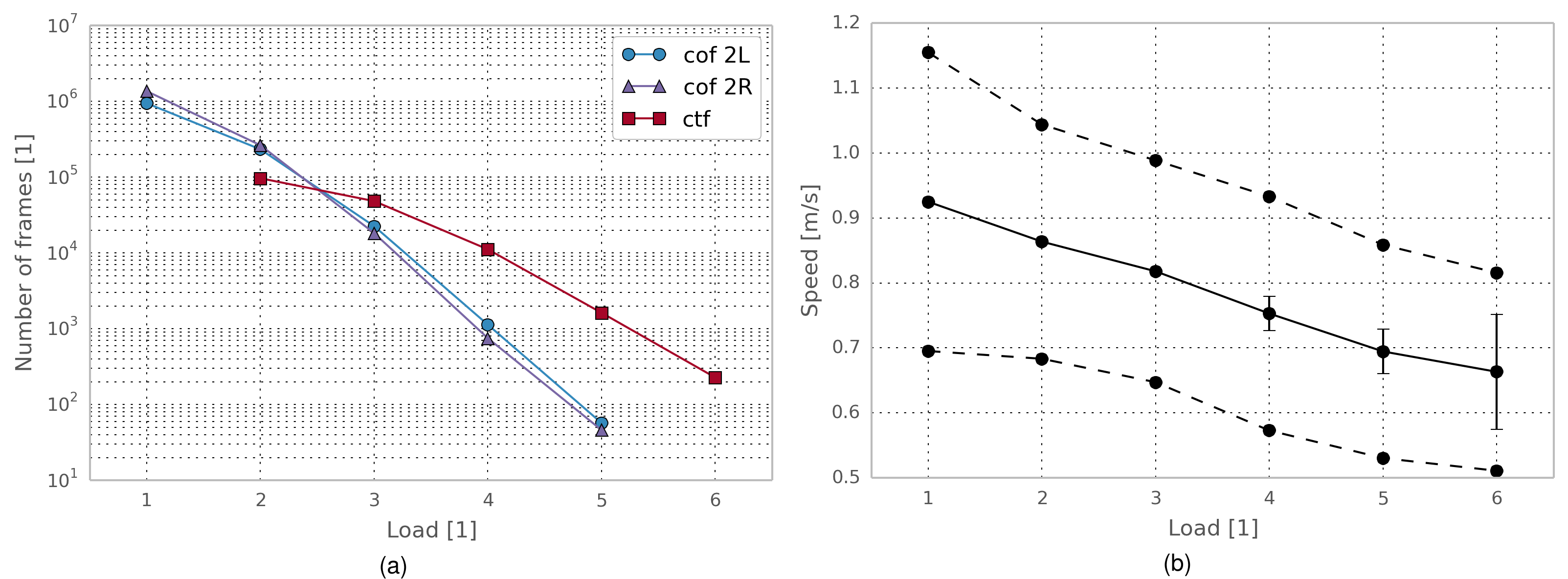}}%
\caption{(a) Absolute frequency of load (amount of people in the facility) depending on the  flow condition (co-flow of all pedestrians to the left (cof 2L) or to the right (cof 2R) or in presence of counter-flowing directions (ctf)). (b) Fundamental diagram: the ensemble-averaged frame-wise mean pedestrian speed  is plotted in dependence on the local facility load, i.e., the amount of people in the facility (solid line). The standard deviation of the average frame-wise speed is reported in added and subtracted to the mean (dotted lines). Confidence on the ensemble mean values decreases as the load increases as high load are less likely to happen; the error bars portray the standard deviation of the ensemble means evaluated after splitting the data-set in four even sub-samples.~\label{fig:load-distrib}}
\end{figure}

In Figure~\ref{fig:load-distrib}(a), we report the load distribution in dependence on the local flow condition. The scenario of one  pedestrian walking undisturbed is the most common, and, in this case, the ascending direction  (left to right)  is the most frequent one. Up to five pedestrians have been observed in co-flow conditions in either directions and with comparable frequencies. Counter-flows appear to be more frequent whenever more than three pedestrians occupy the facility; this is not surprisingly as according to the definition given counter-flow cases include different combinations of the directions for any load greater than two. Up to six pedestrians have been observed in the counter-flow case.

\subsection{Fundamental diagrams \label{s:fdiag}}
In this Subsection we consider a generalized version of the fundamental diagram in which the facility load is considered as a dependent variable (in place of the conventional pedestrian density, by virtue of the reduced area recorded)  and it is compared with the pedestrian velocity (see, e.g.,~\cite{seyfried2008fundamental},~\cite{bellomo2011modeling} and~\cite{venuti2007interpretative} as a reference for conventional fundamental diagrams). Specifically,  we consider the ensemble-mean, frame-wise averaged, pedestrian speed $\bar{U}$ with respect to the instantaneous facility load  and, possibly, flow conditions. In formulas, $\bar{U}$ reads as
\begin{align*}
\bar{U}(&\mbox{ load $= L\ |$ flow condition $= Q$ }) := \\  &mean(\{   \mbox{ $U^n$, for all $n$ such that load = $L$ and flow condition = $Q$ }  \}),
\end{align*} 
where
\[
U^n := mean(\{\mbox{ speed of pedestrians in $f^n$ } \}),
\]
and $Q$ can be, e.g., the single pedestrian case,  the co-flow case, the counter-flow case and so on.

In other words, all the time instants (frames) featuring a given load and flow conditions are grouped, then, the average frame-wise  walking speed, $U^n$, is evaluated as an indicator of the ``effective'' walking speed in the frame. Finally, to have an ensemble indicator, we extract ensemble means of such effective walking speeds $\bar{U}$.

In Figure~\ref{fig:load-distrib}(b), a fundamental diagram including
all possible flow conditions is reported. The diagram, consistently
with other experimental fundamental diagrams (see,
e.g.,~\cite{seyfried2008fundamental}), shows a decreasing trend as the
load increases; moreover it exhibits an approximately linear
behavior. Pedestrians average speed drops from approximately $0.92m/s$
- in the undisturbed case - to approximately $0.68m/s$ - in the most
loaded condition observed. An ensemble standard deviation on the
$\{U^n\}$ decreasing from $0.23m/s$ to $0.15m/s$ can be observed as
the load increases. This fluctuation in the data is likely an effect
the inter-subject and intra-subject variabilities (see, e.g.,
\cite{vzivanovic2007probability}).

\begin{figure}[t]\vspace*{4pt}
\centerline{\includegraphics[scale=.48]{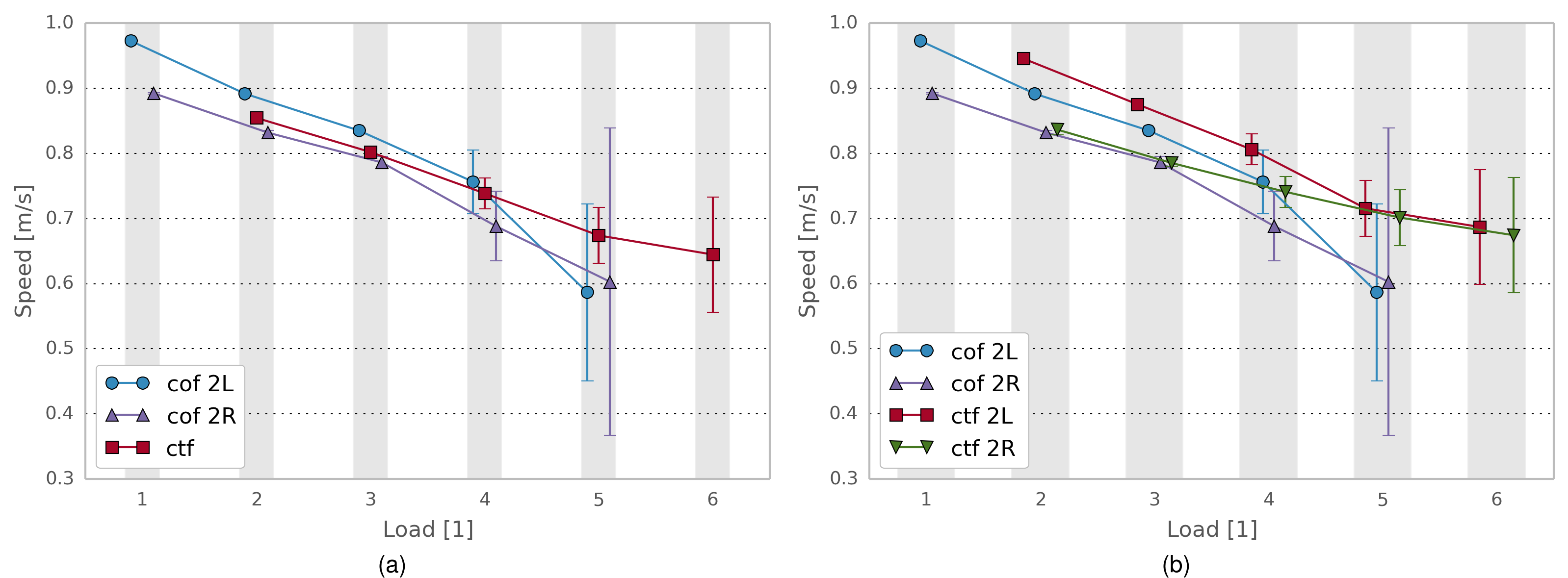}}
\caption{ Fundamental diagrams conditioned to the local flow. The ensemble-averaged frame-wise mean pedestrian speed is plotted in dependence of the local load, in co-flow (all pedestrians have the same direction, cof) condition and in counter-flow condition (at least two pedestrians have different directions, ctf). Both cases of pedestrians going to the left (2L) and to the right (2R) are considered.  (a) Counter-flow cases are considered independently on the direction; (b) counter-flow cases are considered dependently on the direction. Fundamental diagrams have been slightly shifted on the load axis (which can assume only integer values) for enhanced readability. Confidence on the ensemble mean values decreases as the load increases as high load are less likely to happen; the error bars portray the standard deviation of the ensemble means evaluated after splitting the data-set in four even sub-samples. As the condition of co-flow with $load=5$ is very unlikely to happen just few measurements have been collected, hence the large statistical error.~\label{fig:fund-diag-simple}} 
\end{figure}

Qualitative and quantitative changes in the fundamental diagram emerge
when a restriction to the local flow $Q$ is applied. Firstly, it is
worth noting that fundamental diagrams in co-flow conditions (i.e.,
which consider, exclusively, pedestrians going to the left (2L) or
going to the right (2R)) differ one another - see
Fig.~\ref{fig:fund-diag-simple}(a)). This reflects a ``broken
symmetry'' relative to the walking direction. As a matter of facts,
pedestrians moving toward the left side walk faster on average. This
can be explained by considering that the facility is indeed a landing
between two stair cases; pedestrians going to the left have just
descended a ramp of stairs and for this reason they may be walking
slightly faster. On the other hand, pedestrians going to the right
have just climbed a ramp of stairs and for this reason they may be
walking with a lower speed.

When the counter-flow condition is considered, we observe an effective
speed which, for low values of the load, lies in between the two
co-flow cases. This may also be the case at larger loads if one
considers the larger statistics errors present for cases with more
than four pedestrians.

Remarkably if we isolate pedestrians on the basis of their direction
also in counter-flow conditions, we can see that velocities in the
counter-flow cases are always higher or equal than the co-flow cases
with the same load (see Fig.~\ref{fig:fund-diag-simple}(b)). This suggests
that the presence of counter-flows triggers a sort of
self-organization which increases the overall performances in terms of
effective speed.  Similar effects have been observed also
in~\cite{kretz}, where pedestrian fluxes in counter-flows have been
measured to be higher than the fluxes in analogous co-flow conditions.

\subsection{Heat maps \label{s:heat}}
In this Subsection, we analyze how pedestrians (head) positions $\vec{Z}= (X,Y)$ distribute in probability as functions of the local flow conditions. Focus is given on the undisturbed pedestrian case and on two pedestrian counter-flow case.

Probability distribution functions of positions are reported as heat
maps, which, \textit{de facto}, express which portions of the floor
are loaded with higher frequency. Clearly, as the facility is globally
U-shaped, heat maps show a curved trend. This likely reflects an
inertial-like behavior in the act of walking. Moreover, the very shape
of the heat maps appears to be depending both by the walking direction
as well as altered by the local traffic conditions.

\begin{figure}[t]\vspace*{4pt}
\centerline{\includegraphics[scale=.87]{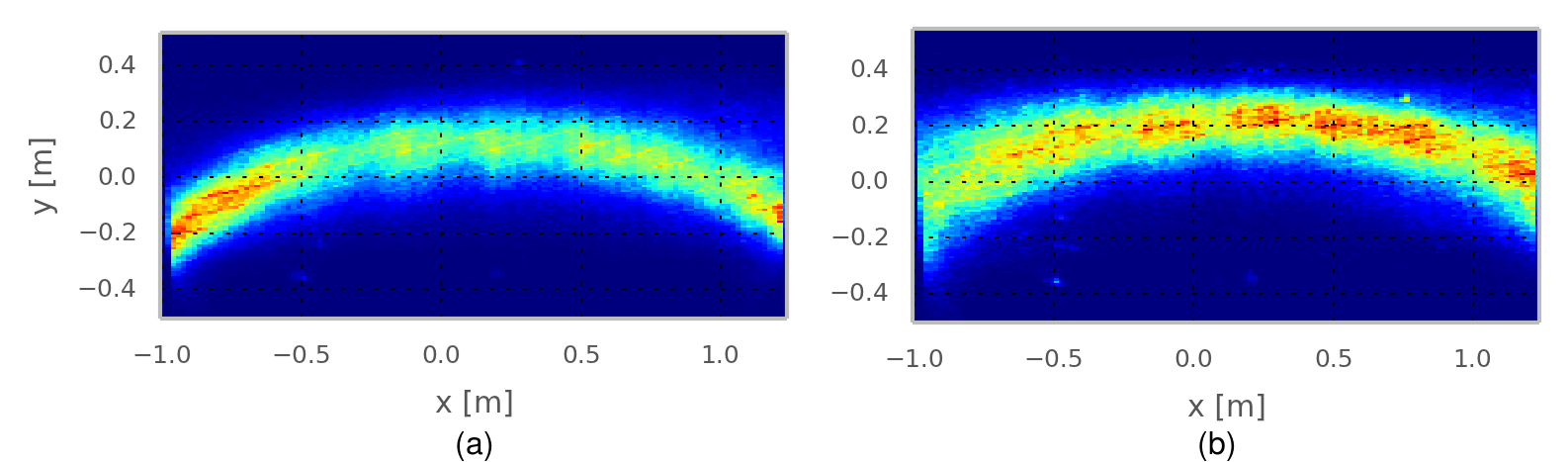}}%
\caption{Heat maps, i.e.~probability distribution functions, of  pedestrians head position. The maps consider  undisturbed pedestrians going to the left (a) and to the right (b). Low probability positions are in blue, high probability positions are in red.~\label{fig:fullheatmap2Reff}}
\end{figure}

In Figure~\ref{fig:fullheatmap2Reff}, the heat maps referring to pedestrians walking undisturbed either from the left to the right side of the facility (Fig.~\ref{fig:fullheatmap2Reff}(a)) or vice-versa (Fig.~\ref{fig:fullheatmap2Reff}(b)), are reported. Pedestrian positions heavily concentrate in a thin layer (say $l = l(x)$) of approximate width of \textit{ca.} $20cm$. 

To compare heat maps with greater ease, 
a \textit{reduced} version of the latter containing just the layer boundaries  is considered. Specifically, for each  chord-wise section  of the facility $x$, local means and local standard deviations of positions,   i.e. $mean(Y | X = x )$ and $std(Y | X = x  )$,  are evaluated. Hence, for each section,  $l = l(x)$ is  approximated as 
\[
l(x) \approx  \left\{y \  \colon \   |y - mean(Y | X = x) | \leq  std(Y | X = x )\right\}.
\]

\begin{figure}[t]\vspace*{4pt}
\centerline{\includegraphics[scale=.87,trim=0cm 0cm 0cm .56cm,clip=True]{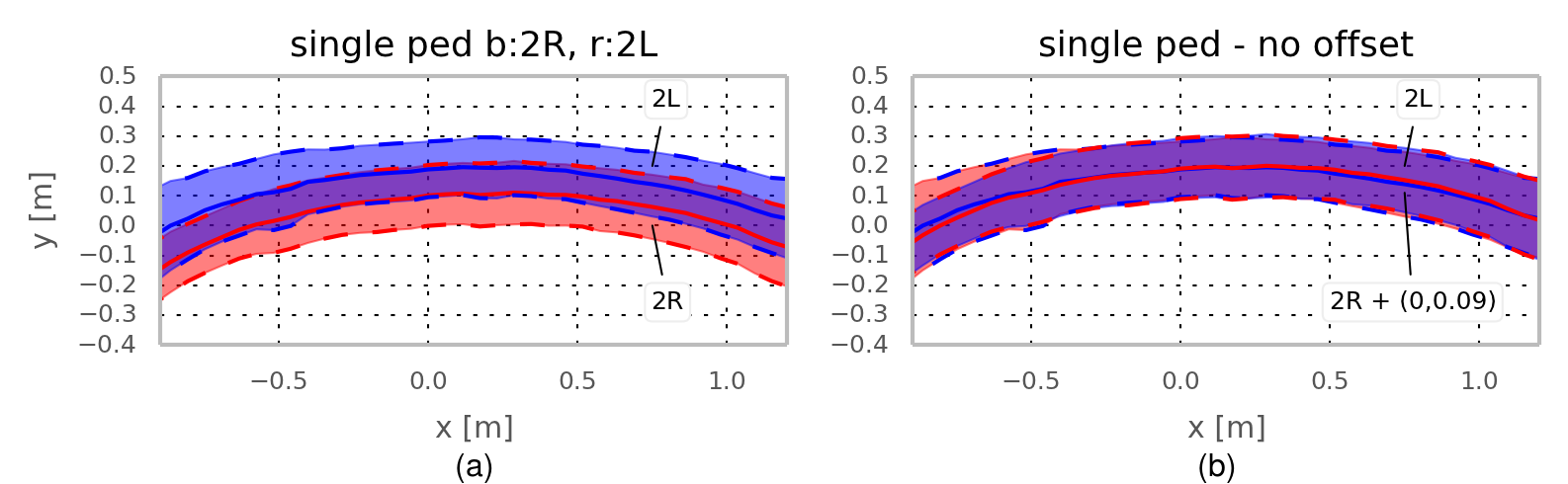}}%
\caption{(a) Reduced heat maps referring to undisturbed pedestrians going left (2L, blue) and right  (2R, red); (b)  heat maps from (a) are superimposed by a vertical shift in upward direction  of the 2R map (red, $\Delta y = .09 m$). The maps feature no qualitative difference than the vertical (chord-wise) translation. ~\label{fig:synth-heatmaps-single}}
\end{figure}

In Figure~\ref{fig:synth-heatmaps-single}(a), reduced heat maps for
single, undisturbed, pedestrians are reported. Pedestrians, even if
free to occupy every region of the corridor, appear to ``naturally''
walk slightly closer to the wall placed at the right hand side of the
walker.  No other qualitative difference in the thin layer geometry is
observed - see Fig.~\ref{fig:synth-heatmaps-single}(b) for a
qualitative comparison.

\begin{figure}[t]\vspace*{4pt}
\centerline{\includegraphics[scale=.87,trim=0cm 0cm 0cm .56cm,clip=True]{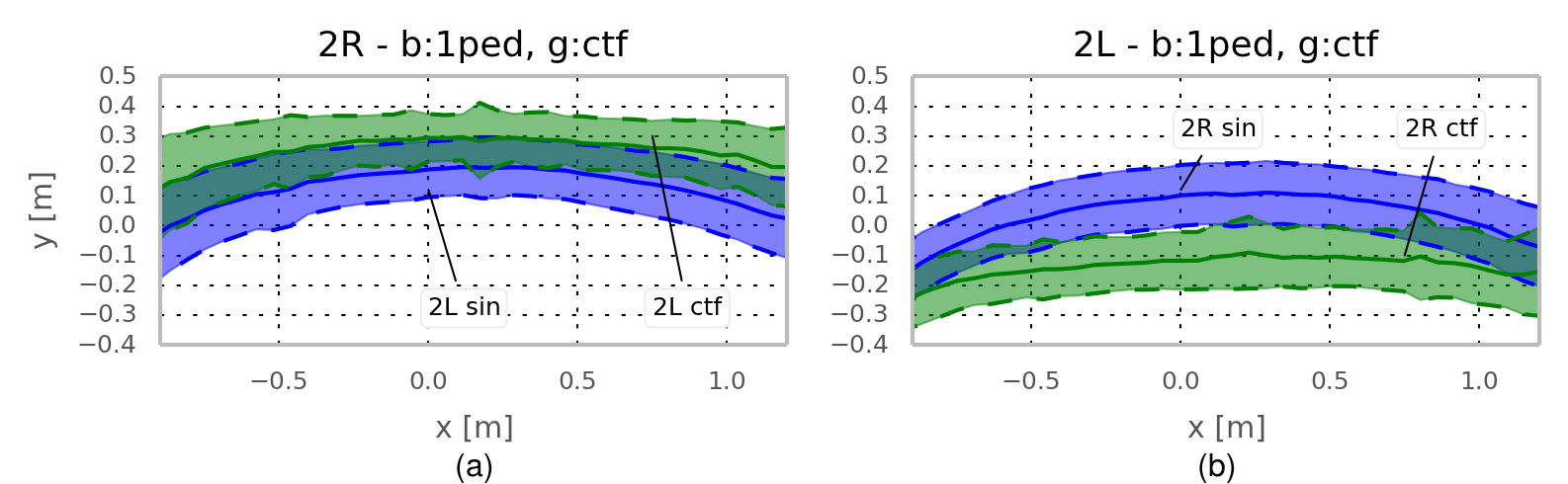}}%
\caption{Heat maps conditioned to the flow. Synthetic heat maps of pedestrians going left (a) and right (b) are compared for the undisturbed case (sin, blue) and the two pedestrians counter-flow case (ctf, green). The vertical shift is determined by the flow condition.~\label{fig:synth-heat-map-pairs}}
\end{figure}

This natural tendency of keeping the right gets heavily emphasized in
presence of a second pedestrians having opposite direction with
respect to the observed one, i.e., in the simplest counter-flow
condition. In Figure~\ref{fig:synth-heat-map-pairs}(a) and (b), the
heat maps of undisturbed pedestrians are compared with the analogous
ones in counter-flow conditions. Pedestrians positions appear to be
pushed to the (relative) right hand sides in close contact with the
wall. In this condition a sort of ``spontaneous organization'' seems
to emerge. This observation is in agreement with laboratory
measurements (cf. e.g. with \cite{kretz}, in which complex
counter-flows have been experimentally induced and analyzed, and with
\cite{moussaid2009experimental}, in which the bias pedestrians have in
choosing mutual avoidance direction has been inquired).

\section{Discussion}\label{Discussion}
In this work a large number of pedestrian trajectories has been
collected and analyzed as ensemble data. The data collection procedure
is grounded on the recording of overhead depth maps - obtained by
means of Microsoft Kinect\texttrademark - which allowed the detection
of pedestrian heads. Head tracking was carried out by Particle
Tracking Velocimetry based techniques.

The data, taken during an experimental campaign lasted fifty working
days, refers to all pedestrians walking on a landing in Eindhoven
University of Technology.  Recorded pedestrians are expected not to be
biased by the recording campaign as no factual modification of the
facility is observable in the ordinary use.
 
The obtained trajectories spread among different flow conditions, from the
single undisturbed pedestrian case, to the co-flow or the counter flow
cases which involve many pedestrians. Respectively, in the co-flow
case every pedestrian walks in the same direction, while, in the
counter-flow cases, at least two opposite directions are observed.

We presented here preliminary results, while our experiment is
collecting additional statistics that will be particularly important
to improve the quality of statistics at higher loads. Trajectories
mostly concentrate at lunch hours, in which the facility load shows to
be homogeneous across the weekdays with the exception of Fridays in
which a $20\%$ reduction of the traffic is noticeable. Most
frequently, just one pedestrian is present in the facility;
nonetheless, up to five pedestrians have been recorded while walking
in co-flow and up to six in counter-flow. As a consequence, very rich
statistics have been collected for loads smaller or equal than four,
but yet not enough for loads corresponding to five or larger number of
pedestrians.

Data has been treated in a statistical fashion including various kind
of averages, possibly conditioned to the flow.  Specifically, ensemble
means of average pedestrians frame-wise speed have been
considered. These allowed us to obtain fundamental diagrams comparing
speeds vs. the local loads. Our fundamental diagrams show that co-flow
speed are higher for descending pedestrians than for ascending ones
(at least up to load four, where statistically significant estimates
of speeds have been obtained). Moreover, speeds in counter-flows
appear to be higher than in corresponding co-flows. This fact may be
related to some kind of spontaneous organization. This organization
may be also noticed in heat maps - i.e., ensemble averages of spatial
positions. A shifting on the average position of pedestrians toward
the right side of the corridor has been measured when counter-flows
occur. The obtained results are likely to feature a dependence on the
specific geometry considered; further experiments on more generic
geometric settings are ongoing.

\section*{Acknowledgments}

We would like to thank A. Holten and G. Oerlemans for the help with
the installation of the Kinect\texttrademark\ sensor in the MetaForum
building at the Eindhoven University of Technology. We thank
A. Liberzon (Tel Aviv, Israel) for his precious help with the adaption
of the Particle Tracking software to our project and A. Tosin (CNR-IAC
Rome, Italy) for the useful discussions. We acknowledge the Brilliant Streets research program of the Intelligent Lighting Institute at the Eindhoven University of Technology.  AC was founded by a “Lagrange” Ph.D. scholarship granted by the CRT Foundation, Turin, Italy and by
the the Eindhoven University of Technology, The Netherlands.

\bibliography{master.bib}
\bibliographystyle{plain}

\end{document}